\documentclass[11pt,a4paper]{article}
\usepackage{epsfig}

\newlength{\dinwidth}
\newlength{\dinmargin}
\newcommand{\Le}{\left(}
\newcommand{\Ra}{\right)}

\newcommand{\beq}{\begin{equation}}
\newcommand{\eeq}{\end{equation}}
\newcommand{\beqar}[1]{\begin{eqnarray}\label{#1}}
\newcommand{\eeqar}{\end{eqnarray}}
\newcommand{\ro}{\varrho}

\begin{document}
\title{\Large \bf 
On the equivalence of Reggeon field theory in zero transverse dimensions 
and reaction-diffusion processes
}
\author{ 
{~}\\
S.~Bondarenko$\,{}^{a)}\,$\thanks{Email: sergb@mail.desy.de}
\hspace{1ex},
L.~Motyka$\,{}^{b),c)}\,$\thanks{E-mail: motyka@th.if.uj.edu.pl}
\hspace{1ex},
A.H.~Mueller$\,{}^{d)}\,$\thanks{E-mail: arb@phys.columbia.edu}
\hspace{1ex},
A.I.~Shoshi$\,{}^{e)}\,$\thanks{E-mail: shoshi@physik.uni-bielefeld.de}
\hspace{1ex} and
B.-W. Xiao$\,{}^{d)}\,$\thanks{E-mail: bowen@phys.columbia.edu}
\\[10mm]
{\it\normalsize ${}^{a)}$ II Institute for  Theoretical Physics, 
University of Hamburg, 22761 Hamburg,  Germany}\\
{\it\normalsize ${}^{b)}$ DESY Theory Group, 22603 Hamburg, Germany }\\
{\it\normalsize ${}^{c)}$ Institute of Physics, Jagellonian University,
30-059 Krak\'{o}w, Poland} \\
{\it\normalsize ${}^{d)}$ Physics Department, Columbia University, 
New York, NY-10027, USA } \\
{\it\normalsize ${}^{e)}$ Fakult\"at f\"ur Physik, Universit\"at Bielefeld, D-33501 Bielefeld, Germany 
}}
\date{September 20, 2006}

\maketitle
\thispagestyle{empty}

\begin{abstract}
The Reggeon field theory in zero transverse dimensions is investigated.
Two versions of the theory are considered: one that allows at most triple
pomeron interactions and the other that embodies an additional~$2\to 2$ 
quartic Reggeon coupling. The behavior of the scattering amplitude at 
asymptotic rapidities is obtained in both cases.
In an $s$-channel picture of the high energy scattering both 
models can be viewed as reaction-diffusion processes. We derive known 
results in Reggeon field theory rather easily using the 
reaction-diffusion formalism. We find that some results which are 
surprising from the Reggeon field theory point of view turn out to have 
a simple interpretation from the reaction-diffusion point of view.
\end{abstract}

\begin{flushright}
\vspace{-19.5cm}
{ BI-TP 2006/36 \\ CU-TP-1161 \\ DESY 06-167}\\
\end{flushright}
\thispagestyle{empty}

\newpage

\section{Introduction}

The high energy scattering in QCD may be described in terms of a 
non-local field theory of QCD pomerons. 
In this formulation the basic degrees 
of freedom --- the BFKL pomerons~\cite{bfkl,bfkl2,bfkl3,bfklsum} --- 
merge and split according to non-local multi-pomeron 
vertices~\cite{Gribov:1984tu,vert1,vert2,eglla,cewerz,braun1,braun2,braun3,bm}. 
This approximation is consistent with the unitarity of the $S$-matrix. 
Alternative approaches to the unitarization of 
high energy amplitudes use extensively methods of statistical physics 
combined with field-theoretical calculations in 
QCD, see e.g.\ Refs.~\cite{Mueller:1993rr}--\cite{stoch8}.
The coordinate space in all those formulations is the transverse
plane to the axis of the high energy collision. In all cases the emerging 
evolution equations for high energy amplitudes are non-local functional
equations which are, so far, prohibitively difficult to solve exactly. 
Therefore, it is useful to study a substantially simplified model 
of the interacting pomeron theory, in which the dependence of the pomeron 
fields on transverse coordinates is discarded. 
The emerging simplified  model -- Reggeon field theory (RFT) 
in zero transverse dimensions -- was proposed and formally solved long 
time ago~\cite{0dimc,0dimq1,0dimq,Amati:1976ck,Ciafaloni:1977xv,0dimi}. 
This toy model exhibits some features which are present also in the more 
realistic QCD pomeron field theory. 
Therefore, RFT in zero transverse dimensions offers a useful testing 
ground for computational methods and recently it enjoys revived 
interest~\cite{Rembiesa:2005gj,Kovner:2005aq,Shoshi:2005pf,Kozlov:2006zj,Shoshi:2006eb,Blaizot:2006wp}. 
In this paper we will explore the equivalence between RFT and the 
statistical reaction-diffusion picture of high energy scattering 
that holds in zero transverse dimensions.

In Section~2 we consider the zero dimensional Reggeon field theory.
Equations of RFT are derived from the Langevin formulation and from the
Lagrangian of RFT.
Two distinct realizations of RFT are studied: a minimal RFT with triple 
pomeron vertices only (which will be abbreviated as MRFT) 
and a theory with an additional $2\to2$ pomeron interaction which is 
equivalent to a simple reaction-diffusion $s$-channel model 
(which will be called the reaction-diffusion RFT or RD-RFT). 
Asymptotic behavior of scattering amplitudes in both RFTs is obtained.
The minimal RFT yields a counter-intuitive
exponential decrease for the scattering amplitude 
$\,T(Y) \sim \exp(-E_0 Y)\,$ at very large rapidities $Y$, 
with the leading order result of $E_0$ given by 
$E_0 \sim \exp(-\mu^2/2\lambda^2)$ where $\mu$ denotes the 
pomeron intercept and $\lambda$ the triple pomeron coupling.
Clearly, the dependence of $E_0$ on the triple pomeron coupling is 
non-perturbative.

The presence of a quartic $2\to2$ pomeron coupling, $\lambda'$, changes this
picture. Namely, since $E_0=0$ for $\lambda'=\lambda^2/\mu$, the scattering 
amplitude approaches a constant value at asymptotic rapidities 
instead of going to zero.
In this case, the exact asymptotic solution of the RFT is given. 
Furthermore, we briefly discuss two possible options of multi-pomeron 
couplings to elementary external sources. Firstly, the eikonal couplings that
 were usually assumed in RFT. In the other scenario only a single pomeron 
coupling to the elementary external source is 
allowed. The latter scheme is 
realized in QCD where (at the leading logarithmic approximation) only single 
BFKL pomeron couples to an elementary colour
dipole~\cite{vert1,vert2,eglla,cewerz}. It is also favored
by the $s$-channel reaction-diffusion picture in which each elementary 
particle in the cascade may couple to at most one $t$-channel exchange.
In this context, it is particularly interesting to consider the RFT counterpart
of the reaction-diffusion model in which only single coupling to external 
sources are allowed with the strength $\alpha\ll 1$ related to the pomeron 
intercept $\mu = \alpha$ and the triple pomeron coupling $\lambda=\alpha^2$. 
In this model the high energy asymptotics, $Y\to \infty$, of the scattering 
amplitude happens to be close to the unitarity limit:
\beq
\tilde T(Y)\, \longrightarrow \, {1\over 1-e^{-1/\alpha^2}},
\eeq 
which should be contrasted with the result of RD-RFT with the same 
parameters and eikonal couplings:
\beq
\label{eq1.2}
T(Y)\, \longrightarrow \, {(1-1/e)^2\over 1-e^{-1/\alpha^2}}.
\eeq

In Section~3 we give a complementary analysis of the scattering amplitudes
in zero transverse dimensions in the probabilistic reaction-diffusion 
framework. 
It turns out, that the minimal RFT may be represented in the $s$-channel 
picture, analogously to the reaction-diffusion process. 
The transition rates, however, do not obey the constraints of the Markovian
process -- the transition coefficients are negative that prohibits 
the genuine probabilistic interpretation. Nevertheless, 
the quasi-reaction-diffusion picture provides an efficient calculational
framework. Thus, we obtain results for high rapidity asymptotics of scattering
amplitudes that coincide with ones from the conventional RFT framework, 
but the interpretation is more transparent. 
We also demonstrate the equivalence of the results obtained within the proper
reaction-diffusion framework with the results of the corresponding RFT.
Finally, the representation of the eikonal model in the reaction-diffusion
framework is given. Curiously, the ``grey disc'' limit of the scattering 
amplitudes given by Eq.~(\ref{eq1.2}) emerges as a consequence of the 
normalisation of the scattering states that is smaller than one.

Finally, we note that a different $s$-channel picture has recently been
elaborated in Ref.~\cite{Blaizot:2006wp} which is very different than the 
one we have here discussed. 
The eikonal couplings are used and boost invariance is 
preserved by having arbitrarily high Reggeon vertices in the hierarchy 
equations. The system does not reach a fixed point, but rather reaches a 
steady rate of increase of particles with increasing rapidity. It will be
interesting to try and further clarify which of these pictures, if either,
has the closest relationship with four-dimensional QCD.

\section{Reggeon field theory in zero transverse dimensions}

\setcounter{equation}{0}
\subsection{Reggeon field theory from the Langevin equation.}

The general form of the Langevin equation in the Ito formulation is
\beq\label{FP1}
du(y+dy)\,=\,D_1(u(y))\,dy\,+\,D_2(u(y))\,d\omega(y)\,,
\eeq
where $u(y)$ is a stochastic function of an evolution variable $y$, 
$dy$ is an infinitesimally small step of $y$, and
$\omega(y)$ is related to a random noise variable $\nu(y)$,
\beq\label{FP3}
d\omega(y)\,=\,\nu(y)\,dy\, .
\eeq
The noise is taken to be Gaussian,
\beq\label{FP6}
\langle \, \nu(y)\, \rangle \,=\,0, \qquad\quad
\langle \,\nu(y)\,\nu(y')\, \rangle\,=\,\delta\,(y-y')\,.
\eeq

Let us consider a dynamical quantity $G(u)$ with $u$ distributed  
with the probability density $p(y,u)$. It follows from the Ito calculus 
that the average value of this variable
\beq\label{FP8}
\langle\, G \,\rangle_y\,=\,\int du\,p(y,u)\,G(u)\,,
\eeq
evolves according to the following master equation:
\beq\label{FP7}
\frac{d\langle\,G\,\rangle_y }{dy}\,=\,
\left\langle D_1(u)\,\frac{dG}{du}\,\right\rangle_y\,+\, 
\left\langle\, \frac{D_2^2(u)}{2}\,\frac{d^2G(u)}{du^2}\,\right\rangle_y\,.
\eeq
Thus, assuming that moments of the variable $u$, 
\beq\label{FP9}
\langle\, u^k\, \rangle_y\,=\,\int du\,p(y,u)\,u^k\,
\eeq
obey the hierarchy of (deterministic) equations
\begin{eqnarray}\label{FP10}
\frac{d\langle\,u^k\,\rangle_y}{dy}\,&\,=\,&\, 
k\, \left\langle\, D_1(u) u^{k-1} \,\right\rangle_y \, + \, k(k-1)\,
\left\langle\,\frac{D^2_2(u)}{2} u^{k-2}  \,\right\rangle_y.
\end{eqnarray}
which closes if $D_1(u)$ and $D^2_2(u)$ are analytic functions of $u$.
With an appropriate choice of the functions $D_1(u)$ and $D_2(u)$ 
this hierarchy of equations may be made equivalent to the 
equations of the Reggeon field theory in zero transverse dimensions. 
We choose the following form of $D_1(u)$ and $D_2(u)$ 
\begin{eqnarray}\label{FP11a}
D_1(u)\,& = & \,\alpha\,u\,-\,\beta\,u^2\, ,\\
D^2_2(u)\,& = & \,2\,(\alpha'\,u\,-\,\beta'\,u^2\,)
\label{FP11b} ,
\end{eqnarray}
which is equivalent to RFT with triple pomeron vertices and a quartic 
$2\to 2$ pomeron vertex.
Indeed, with (\ref{FP11a}) and (\ref{FP11b}), one obtains from (\ref{FP10}) 
a closed set of equations
\beq\label{FP12}
\frac{d\langle\,u^k\,\rangle_y}{dy}\,=\,\alpha\,k\,\langle\,u^k\,\rangle_y\,+\,
\alpha'\,k\,(k-1)\,\langle\,u^{k-1}\,\rangle_y
\,-\,\beta\,k\,\langle\,u^{k+1}\,\rangle_y
\,-\,\beta'\,k\,(k-1)\,\langle\,u^{k}\,\rangle_y.
\eeq
As it will be shown, those equations are the same as equations of RFT if 
the $k$-pomeron amplitude $\Phi_k(y)$ is connected to the $k$-th moment
$\langle\,u^{k}\,\rangle_y$ in the following way,
\beq\label{npsi}
\Phi_k(y) \,= \, -(-\alpha)^k\, \frac{ \langle\,u^{k}\,\rangle_y}{k!},
\eeq
and the parameters $\alpha'$, $\beta$ and $\beta'$ are suitably adjusted.

\subsection{Hamiltonian formulation of RFT}

The analysis in this and the following parts of this section 
recapitulates the formulation and the key results that were obtained within
RFT long ago~\cite{0dimq1,0dimq,Amati:1976ck,Ciafaloni:1977xv,0dimi}. 
We describe them in some details in order to facilitate a comparison
with the following analysis within the $s$-channel framework.

Originally, RFT was formulated as a field theory 
(or as quantum mechanics in zero transverse dimensions) of pomerons.
The basic degrees of freedom in this formulation are the Gribov
fields $\psi$ and  $\psi^{+}$ that create and annihilate the pomeron.
The action defining the theory with triple pomeron couplings only (MRFT) 
is defined in the following way:
\beq\label{Lag1}
S \,=\, \,\int dy \, \left\{ 
\psi^{+}\,\partial_y {\psi}\,-\,\mu\,\psi^{+}\,\psi\,+\,i\,
\lambda\,\psi^{+}\,(\psi^{+}\,+\,\psi)\,\psi\,\,\,\right\},
\eeq 
where $\,\mu\,$ is the bare intercept of the pomeron and
$\lambda$ is the coupling of the triple pomeron interaction.
Note that the action is symmetric under the transform
\beq
\psi \,\longleftrightarrow\, \psi^+, \qquad y\,\to\, -y,
\eeq 
which corresponds to the symmetry between the target and the projectile.
After redefinition of the Gribov fields, $q\,=\,i\,\psi^{+}\,$ and
$p\,=\,i\,\psi\,$ the action may be rewritten in terms of real 
quantities
\beq\label{Lag2}
S \,=\, \int dy \, \left\{ 
p\, \partial_y q \,+\,\mu\,q\,p\,-\,\lambda\,q\,(q\,+\,p)\,p\,\
\right\}.
\eeq 
Then, the Hamiltonian of the problem is given by
\beq\label{Ham1}
H=\,\mu\,q\,p\,-\,\lambda\,q\,(q\,+\,p)\,p\,\,\,.
\eeq
The commutation relation of the pomeron annihilation and creation operators,
$[\psi,\psi^+]=1$, implies $[p,q]=-1$ which may be realized for instance 
by identification of $q$ with a position operator and $p$ with a differential
operator, 
\beq\label{Oper}
\,p\,=\,-\frac{\partial}{\partial\,q}\,.
\eeq
In this representation the Hamiltonian takes the form
\beq\label{Ham2}
H \,=\,-(\mu\,q\,-\,\lambda\,q^{2})\,\frac{\partial}{\partial q}
\,-\lambda\,q\,\frac{\partial^2}{\partial q^2}\,.
\eeq
In the following the ratio $\ro \,=\,\mu\,/\,\lambda\,$ will be used, 
which is the large parameter of our model. 

The interacting pomeron system whose state is given by 
$\Psi(y,q)$ evolves according to a Schr\"{o}dinger equation 
(with an imaginary evolution variable)
\beq\label{Ham3}
- \frac{\partial\,\Psi}{\partial y}\,=\,H\,\Psi.
\eeq 
The complete definition of the problem requires that the domain of 
$q$ and the boundary conditions are specified. 
This, however, depends on the particular choice of the couplings. 
For the minimal RFT it was shown that it is sufficient to consider
$0 \leq q \leq \infty$ with $\Psi(y,q)$ that obeys the conditions
\beq
\label{H4}
\Psi(y,0)=0, \qquad \lim_{q\to\infty } \partial_q \Psi(y,q) = 0
\eeq
at all rapidities $y$.

The state $\Psi(y,q)$ embodies the complete information about
$k$-pomeron amplitudes $\Phi_k(y)$ in the $t$-channel,
\beq \label{Quant14}
\Psi(y,q)\,=\,\sum_{k=1}^{\infty}\,\Phi_{k}(y)\,q^k .
\eeq
The following hierarchy of equations, equivalent to (\ref{Ham3}),
\beq\label{Quant15}
\frac{d\Phi_{k}}{dy}\,=\,\mu\,k\,\Phi_{k}\,+\,\lambda\,k\,(k+1)\,
\Phi_{k+1}\,-\,\lambda\,(k-1)\,\Phi_{k-1}\,, \qquad k=1,2,\ldots
\eeq
is identical to the hierarchy of equations (\ref{FP12}) describing the 
evolution of moments $\langle u^k \rangle$ in the Langevin 
formulation after the substitution (\ref{npsi}) provided that it is set
\beq
\alpha = \mu, \qquad \alpha\alpha' = \beta/\alpha  = \lambda, 
\qquad \beta' = 0.
\eeq
Note, that the target-projectile symmetry of RFT imposes the constraint
\beq
\beta =  \alpha^2\alpha'
\eeq
on the Langevin formulation.

\subsection{External couplings and scattering amplitudes}

The scattering amplitudes in the Reggeon field theory may be obtained after the
bulk action given by Eq.~(\ref{Lag1}) is supplemented by the action describing
the coupling of the pomeron system to external particles. In the original 
formulation of RFT the eikonal couplings were assumed, that is the 
coupling of $k$ pomerons to the external source was given by a product of $k$ 
single pomeron couplings multiplied by the combinatoric symmetry factor, $1/k!$
for $k$ outgoing pomerons
\footnote{This assumption seems to be rather natural for 
interacting point-like objects, but it may be not 
applicable in the case of QCD pomerons. In particular, due to a 
complex non-local nature of the BFKL pomeron, only one BFKL pomeron can 
couple to an elementary colour dipole in the leading logarithmic 
approximation.}.  
In this scenario, the initial condition of the RFT Hamiltonian evolution 
is obtained by a resummation of multi-pomeron source amplitudes:
\beq
\label{ext1}
\Psi(y=0,q)\,=\, \sum_{k=1} ^{\infty} \Phi_k(0) q^k \,=\, \sum_{k=1} ^{\infty} { -(-g_1) ^k q^k \over k!} \, = \, 1-\exp(-g_1 q), 
\eeq
where $g_1$ is the coupling of the pomeron to the source. 
Thus, the scattering amplitude in RFT at rapidity $y$ is given by
\beq
\label{ext2}
T(Y;g_1,g_2) \,=\, \sum_{k=1} ^{\infty} \Phi_k(Y) g_2^k  = \Psi(Y,g_2),
\eeq
where $g_2$ is the pomeron coupling to the projectile.

In the alternative realization of RFT only a single pomeron coupling is possible to an elementary (thus not evolved) external source. This assumption naturally emerges if one considers dipole-dipole scattering in QCD. Then, the initial condition of the Hamiltonian evolution is given by:
\beq
\label{ext3}
\tilde{\Psi}(y=0,q)\,=\, \Phi_1(0) q \,=\, g_1 q,
\eeq
and the scattering amplitude may be obtained as
\beq
\label{ext4}
\tilde{T}(Y;g_1,g_2) \,=\, g_2\, \Phi_1(Y)  \, = \, g_2\,
\left. {\partial \Psi(Y,q) \over \partial q}\right|_{q=0}.
\eeq

\subsection{High rapidity asymptotics of the minimal RFT}

The solution to the Hamiltonian problem (\ref{Ham3}) may be
given in terms of an infinite summation over a discrete set of
eigenstates with wave function $\psi_n(q)$
\beq\label{Ham4}
\Psi(y,q)\,=\,\sum_{n=0}^{\infty}\,\lambda_{n}\,
e^{-E_{n}\,y}\,\psi_{n}(q)\,\,,
\eeq
with the eigenvalues of the Hamiltonian $H$ denoted by $E_{n}$,
$n=0,1,2,\ldots$. Clearly, the asymptotic rapidity dependence of the 
scattering amplitude is governed by the lowest eigenvalue $E_0$. 

Minimal RFT is known to exhibit a counter-intuitive feature of an asymptotic
decrease of the scattering amplitude down to zero at $Y \to \infty$.
This happens because the lowest eigenvalue $E_0$ of the RFT Hamiltonian is
positive. We shall recall a heuristic derivation of $E_0$, given in
Ref.~\cite{Ciafaloni:1977xv} that demonstrates this explicitely.

First, let us transform the eigenfunction corresponding to $E_0$:
\beq\label{Quant3}
\psi_{0}(q)\,=\,e^{(q-\ro)^2\,/4}\,f_{0}(q)\,\,,
\eeq 
that leads to the following equation for $\,f_{0}(q)\,$:
\beq\label{Quant4}
H' \,f_{0}(q)\,=\,
\Le
-\frac{\partial^2\,}{\partial q^2}\,-\,\frac{1}{2}\,+\,
\frac{1}{4}\,(q\,-\,\ro)^2\,\Ra\,f_{0}(q)
\,=\, \frac{E_0}{\lambda\,q}\,f_{0}(q)\,,
\eeq
and the boundary conditions in Eq.~(\ref{H4}) impose the following behavior of 
$f_{0}(q)$:
\begin{eqnarray}\label{Quant5}
\,&\,&\,f_0(q= 0)= 0\,\\
\,&\,&\,f_0(q\,\rightarrow\,\infty)\,\propto \,
e^{-q^2\,/\,4\,+\,q\ro\,/\,2}\,\,.
\end{eqnarray}
This eigenvalue problem may be solved approximately starting from a simpler
problem of the ground state of the harmonic oscillator, with only modification 
due the initial condition Eq.~(\ref{Quant5}). 
In this case, the {\em Coulomb term} $E_0/\lambda q$ is treated as a 
small perturbation. Then the value $E_0$ must be adjusted so that the
lowest eigenvalue of the operator $\,H'-E_0/\lambda q\,$ vanishes. 
Fortunately, the eigenfunctions of operator $H'$ in Eq.~(\ref{Quant4})
defined on $L_{R}^{2}(0,\,\infty)$ are known. 
These are parabolic cylinder functions $D_{\nu}(q-\ro)$:
\beq\label{Quant6}
H'\,D_{\nu}(q-\ro)\,=\,\nu\,D_{\nu}(q-\ro)\,.
\eeq
It emerges self-consistently that $\nu$ 
quickly vanishes with increasing $\ro$, specifically 
$\nu\,\sim e^{-\ro^2\,/2}\,$. 
The initial condition at $q\,\rightarrow\,0$ implies
\beq\label{Quant7}
D_{\nu}(-\ro)\,=\,0\,,
\eeq
fulfilled by $\nu=\nu_0$ that may be obtained approximately after 
performing asymptotic expansion of $D_{\nu}(-\ro)$ in terms of the 
large parameter $\ro$: 
\beq\label{Quant8}
\nu_{0}\,\simeq\,\frac{\ro}{\sqrt{2\pi}}\,e^{-\ro^2\,/2}\,(1+O(1/\ro^2)).
\eeq
 
In the first order of the perturbative analysis in $E_0$, the contribution
of the {\em Coulomb term} is given by
\beq\label{Quant9}
\left\langle {E_{0} \over \lambda q}\right\rangle
\,\simeq\,
\frac{E_{0}}{\lambda}\,\int\,
\frac{D_{\nu_0}^{2}(q-\ro)}{q}\,dq\,\simeq \,
\,\frac{E_{0}}{\lambda\,\ro}\,(1+O(1/\ro^2))\,.
\eeq
This correction must be adjusted so that it cancels to zero the
energy of the operator $\,H'-E_0/\lambda q\,$. Thus it follows from
Eq.~(\ref{Quant6}) and  Eq.~(\ref{Quant9}) that
\beq\label{Quant10}
\,\frac{E_{0}}{\mu}\,=\,\nu_{0},
\eeq
leading to the energy of the ground state
\beq\label{Quant11}
E_{0}\,\simeq\, \frac{\mu\,\ro}{\sqrt{2\pi}}\,
\,e^{-\ro^2\,/2}\,(1+O(1/\ro^2))\,,
\eeq
and the behavior of the RFT wave function at large rapidities,  
\beq\label{Quant12}
\Psi(y,q)\,\sim\,e^{(q-\ro)^2\,/4}\,D_{\nu_0}(q-\ro)\,
e^{-\frac{y}{\sqrt{2\pi}}\,\mu\,\ro\,
\,e^{-\ro^2\,/2}\,}\,.
\eeq
Therefore, the scattering amplitudes at fixed external couplings
decrease exponentially with increasing rapidity
\beq\label{Quant13}
T(Y,g_1,g_2)\,\sim \,\exp\left[-\frac{Y}{\sqrt{2\pi}}\,\mu\,\ro\,
\,e^{-\ro^2\,/2}\,\right].
\eeq
Note, that the exponent $E_0$ is exponentially small at large $\ro$,  
$E_0 \sim e^{-\ro^2}$, and that the splitting of $E_0$ from zero is
a non-perturbative effect in the triple pomeron coupling, 
$\lambda = \mu / \ro$. This exponential decrease 
of the amplitude may be interpreted as a result of a tunneling phenomenon~
\cite{0dimq1,Ciafaloni:1977xv,0dimi}.
In fact, the tunneling rate was explicitly calculated using the instanton 
calculus~\cite{0dimi} which gives exactly the $Y$-dependence given by 
Eq.~(\ref{Quant13}). Thus, the semi-classical calculation of the (eikonal) 
MRFT scattering amplitude at large~$Y$ yields:
\beq
\label{Q14}
T(Y,g_1,g_2)\,\simeq\, 
(1-e^{-\ro g_1})\,(1-e^{-\ro g_2})\,
\exp\left[-\frac{Y}{\sqrt{2\pi}}\,\mu\,\ro\,\,e^{-\ro^2\,/2}\,\right].
\eeq

\subsection{Markovian RFT}

Another special case of RFT is obtained for the choice
\beq
\alpha' = \alpha, \qquad \beta' = \beta = \alpha^3,
\eeq 
in the Langevin equation. In this case the evolution of the system may 
be represented in terms of a Markovian reaction-diffusion process 
in $0$-dimensions with the $1\to 2$ particle splitting probability $\alpha$ 
and $2\to 1$ particle merging probability $\beta= \alpha^3$. 
To be precise, in this case hierarchy of evolution equations (\ref{FP12}) 
for moments $\langle u^k \rangle$,
\beq\label{mark1}
\frac{d\langle\,u^k\,\rangle_y}{dy}\,=\,\alpha\,k\,\langle\,u^k\,\rangle_y\,+\,
\alpha\,k\,(k-1)\,\langle\,u^{k-1}\,\rangle_y
\,-\,\alpha^3\,k\,\langle\,u^{k+1}\,\rangle_y
\,-\,\alpha^3\,k\,(k-1)\,\langle\,u^{k}\,\rangle_y,
\eeq
is identical to the hierarchy of evolution 
equations (\ref{eq3.11}) for factorial moments $n^{(k)}$ 
of particle number in the reaction-diffusion process. 
This version of RFT, which we denote as RD-RFT, is defined by 
the following hierarchy of evolution equations for $k$-pomeron amplitudes
\beq\label{mark2}
\frac{d\Phi_{k}}{dy}\,=\,\mu\,k\,\Phi_{k}\,+\,\lambda\,k\,(k+1)\,
\Phi_{k+1}\,-\,\lambda\,(k-1)\,\Phi_{k-1}\,-\,\lambda'\,k\,(k+1)\,\Phi_{k}\,
\eeq 
with the pomeron intercept $\mu=\alpha$, the triple pomeron coupling
$\lambda=\alpha^2$ and the quartic $2\to 2$ pomeron coupling 
$\lambda'=\alpha^3$. Consequently, the Hamiltonian of the system reads
\beq\label{FV2}
H_M\,=\,(\lambda\,q^{2}\,-\,\mu q)\,\frac{\partial}{\partial q}\,+
\,(\lambda'\,q^2\,-\,\lambda\,q)\,\frac{\partial^2}{\partial q^2}.
\eeq
In general, the Hamiltonian $H_M$ factorizes into a $q$-dependent part 
(without derivatives) and a $\partial/\partial q$-dependent part
provided that $\lambda'/\mu = (\lambda/\mu)^2$. This condition is obeyed
in particular by values of  $\lambda$ and $\lambda'$ imposed by the relation 
to the reaction-diffusion model and the Hamiltonian reads
\beq
H_M \, = \,
\alpha \, \left( \alpha q^2 - q\right) \,
\left(\frac{\partial}{\partial q} + 
\alpha\frac{\partial^2}{\partial q^2} \right).
\eeq
In order to find the scattering amplitude we consider the quantum mechanical 
problem defined by this Hamiltonian:
\beq\label{FV3}
-\frac{\partial\,\Psi}{\partial\,y} \,=\,H_M \Psi\,.
\eeq
The boundary conditions that should be imposed on $\Psi(y,q)$ in the 
regular singular points $q=0$ and $q=\ro=1/\alpha$ are the following:
\begin{eqnarray}\label{QFV2}
\,&\,&\,\Psi(y,q=0)\,=\,0\\
\,&\,&\,\Psi(y,q) \;\mbox{ is an analytic function around }q=1/\alpha\, .
\end{eqnarray}
The solution may be represented in terms of the eigenfunctions $\psi_n(q)$ and
eigenvalues of the Hamiltonian $H_M$ 
\beq\label{FV4}
\Psi(y,q)\,=\,\sum_{n=0}^{n=\infty}\,\lambda_{n}\,
e^{-E_{n}\,y}\,\psi_{n}(q)\,.
\eeq
Clearly, the state characterized by
\begin{eqnarray}\label{QFV3}
\,&\,&\,E_0\,=\,0\,\\
\,&\,&\,\psi_{0}(q)\,=\,1\,-\,e^{-q/\alpha}\,,
\end{eqnarray}
is an eigenstate. In fact, this is the ground state of $H_M$ which dominates
the scattering amplitude at high rapidities $y$. All the higher states
with $n=1,2,\ldots$ obey the condition\footnote{
This may be easily seen from an analysis of the power-series solutions
of the eigenvalue equation, $\,H_M \psi_n = E_n \psi_n\,$, 
around the regular singular point $q=\ro$. 
Namely, for any $E_n \neq 0$ there exists an analytic 
solution which approaches zero as $\,q-\ro\,$ at $q \to \ro$. 
The other independent solution does not vanish at $q=\ro$ but it 
contains non-analytic  terms $\, \sim \log (\ro-q)\,$ and thus 
it is rejected.}
$\psi_n(\ro) = 0$, see e.g.\ Ref.~\cite{Kozlov:2006zj}.   
Therefore, the weight $\lambda_{0}$ in Eq.~(\ref{FV4}) may be found 
from the initial condition at $y=0$,
\beq\label{QFV4}
\Psi(y=0,q=1/\alpha)\, = \, \lambda_{0}\,\psi_{0}(1/\alpha)\,=\, 
\lambda_0\, (1\,-\,e^{-1/\alpha^2}\,),
\eeq
and the solution at large $y$ is approximately given by 
\beq\label{CVF7}
\Psi(y,q)\,\simeq\, {\Psi(0,1/\alpha) \over 1-e^{-1/\alpha^2}}\, (1-e^{-q/\alpha}).
\eeq

Assuming the eikonal couplings to the external sources, characterized by coupling constants $g_1$ and $g_2$, the initial condition takes the form 
$\Psi(0,q)=1-\exp(-g_1 q)$, and the scattering amplitude at high~$Y$ reads
\beq\label{CVF9}
T(Y,g_1,g_2)\,\simeq\,
\,\frac{(1\,-\,e^{-\,g_1/\alpha})\,(1\,-\,e^{-g_2/\alpha})}
{\,1\,-\,e^{-1/\alpha^2}\,}\,.
\eeq
For $g_1=g_2= \alpha \ll 1$ and $Y\to \infty$ 
this amplitude tends to $(1-1/e)^2 /(1-e^{-1/\alpha^2})$ 
which is far from the unitarity limit.

In the standard reaction-diffusion picture only coupling of single pomeron to
the elementary external source is allowed, with the strength $\alpha$,
and $g_1=g_2=\alpha$. Then, with the initial condition $\Psi(0,q)=q/\alpha$, 
one obtains for $\alpha Y \gg 1$ 
\beq\label{CVF11}
\left. \tilde{T}(Y,g_1,g_2) \right|_{ g_1=g_2=\alpha}
 \;\simeq\; 
\left.
{(g_1/\alpha)\, (g_2/\alpha) \over 1-e^{-1/\alpha^2}}
\right|_{ g_1=g_2=\alpha}
\;=\; {1\over 1-e^{-1/\alpha^2}}. 
\eeq

As we will see in the next section, the results for the amplitude 
given by Eq.~(\ref{CVF9}) or Eq.~(\ref{CVF11}) may be reproduced in the 
reaction-diffusion model confirming the equivalence of the approaches.


\section{ High energy behavior of the scattering amplitude in 
diffusion reaction framework}
\setcounter{equation}{0}

\indent In this section we use the reaction-diffusion formalism to 
reproduce the above results in Reggeon field theory.  This yields two 
benefits.  First the results of Reggeon field theory emerge rather 
easily using the reaction-diffusion formalism and, secondly, some 
seemingly surprising results from the Reggeon field theory point of 
view have a simple interpretation from the reaction-diffusion point 
of view.

\subsection{$3P$ coupling only}

\indent High energy scattering where only a three pomeron interaction 
is present in the Reggeon field theory is a very natural theory to 
study from the $t$-channel (Reggeon field theory) perspective.  It 
does not naturally correspond to a reaction-diffusion theory but, 
formally, one can force it into such a description.  As we have seen 
in the previous section the  hierarchy equation for the factorial 
moments, $n^{(k)}$, is

\beq
{dn^{(k)}\over dy} \, = \, \alpha kn^{(k)}+\alpha k (k-1) n^{(k-1)}-\beta 
k n^{(k-1)}
\label{eq3.1}
\eeq

\noindent with

\beq
\label{eq3.2}
n^{(k)}(y)\,=\,\sum_{n=k}^\infty P_n(y)\, n(n-1)\cdot\cdot\cdot(n-k+1)
\eeq

\noindent and where $P_n$ is the probability of having exactly  $n$ 
particles in the system at time (rapidity) $y.$  The hierarchy 
equation is equivalent to the master equation

\beq
\label{eq3.3}
{dP_n\over dy} \,=\,  - \alpha n P_n + \alpha (n-1) P_{n-1} + \beta 
(n+1)(n+2) P_{n+2}-\beta n (n+1) P_{n+1}
\eeq

\noindent which requires an elementary $1 \to 2$ splitting with 
transition rate $\alpha,$ a $2 \to 0$ vertex having transition rate 
$\beta$ and a $2\to 1$ vertex with transition rate $-\beta.$  These 
later two vertices are certainly not natural in a genuine 
reaction-diffusion process, especially the $2\to 1$ vertex with 
negative probability.  Nevertheless, formally, such vertices in a 
reaction-diffusion context are equivalent to a Reggeon field theory 
having only a triple pomeron coupling.  The presence of a $2\to 1$ 
vertex with a negative transition rate means that $P_n$ is not 
guaranteed to be positive but, as we shall see, this does not appear 
to cause difficulties in the situation we shall be considering.

\indent The presence of a $2\to 0$ vertex means that

\beq
\label{eq3.4}
{dP_0\over dy} \,=\, 2\beta P_2
\eeq

\noindent so that there are transitions into the zero particle state. 
However, there are no transitions out of the zero particle state, as 
is apparent from (\ref{eq3.3}), so that ultimately the system will settle 
down completely into the zero particle sector.  However, in the 
situation of interest to us, $\beta = \alpha^3$ is very small so that 
$P_2$ at early times is very small and thus the lifetime of the 
system is very long, of size $\,e^{1/2\alpha^2}\,$ as we shall find 
below.  Thus the system will first settle down, rather quickly, into 
a quasi-stable state and then \underline{slowly} all the probability 
will appear in the $n=0$ sector.  This slow transition to the $n=0$ 
state will correspond to the tunneling of the $3P$ Reggeon field 
theory discovered long ago.

To be specific suppose at $y=0$ we take

\beq
\label{eq3.5}
P_n(0) \,=\, \delta_{n 1},
\eeq

\noindent that is we start with a single particle.  So long as 
$\alpha/\beta \gg 1$ probability rapidly flows to higher  $n$ states, 
according to (\ref{eq3.3}), via $1\to 2$ transitions.  When  $n$ reaches size 
$\alpha/\beta$ the $2\to 0$ and $2\to 1$ transitions become important 
and the system becomes metastable.  It is easy to check that

\beq
\label{eq3.6}
P_n \,=\, {c\over 2}\left(
{\alpha\over \beta}
\right)^{n+1\over 2}{(n-2)!!\over n!} 
e^{-\alpha/(2\beta)}
\eeq

\noindent with $(-1)!!\equiv 1,$ gives ${dP_n\over dy}=0$ for $n > 1$ 
when (\ref{eq3.6}) is substituted into (\ref{eq3.3}) and furthermore 
that this is the case even when a different value of  $c$  is taken for even 
values of $n$ as compared to odd values of $n.$ The $n=0$ part of 
(\ref{eq3.3}) is given by (\ref{eq3.4}) which, along with the $n=1$~part 
of (\ref{eq3.3}),
is the only departure from (\ref{eq3.6}) being a fixed point solution.  
Now starting from (\ref{eq3.5}) we expect that $P_n$ 
should be smooth in $n$ at large values of $n.$  In order to insure 
this smoothness one must choose the $c$ in (\ref{eq3.6}) to be different for 
even and odd values of $n.$  A simple calculation shows that if one 
takes

\begin{displaymath}
c\,=\,1\ \ \ \ \ \ \ \ \ \ \mbox{for odd $n$,}
\end{displaymath}
\beq
\label{eq3.7}
c \,=\, {\sqrt{2/\pi}}\ \ \ \ \ \ \ \ \mbox{for even $n$}
\eeq

\noindent then (\ref{eq3.6}) is smooth in $n$, for large $n$, and $P_n$ is a 
normalized probability distribution when $\alpha/\beta \gg 1$.  One 
can expect the form (\ref{eq3.6}) to set in at a ``time''

\beq
\label{eq3.8}
y_0\, \simeq \, 1/2\beta 
\eeq

\noindent after the start of the evolution of the system from the 
initial distribution (\ref{eq3.5}).

\indent Now as time goes by more and more of the probability of the 
system will be located in the $n=0$ state.  This means that one must 
change the normalization of (\ref{eq3.6}) to account for the loss of 
probability in the normal $n \not= 0$ states.  This is easily done by 
including a factor $1-P_0(y)$ on the right hand side of (\ref{eq3.6}).  
One can then, in general, write (\ref{eq3.4}) as

\beq
\label{eq3.9}
{d\over dy} (1-P_0) \, = \, - 2\beta P_2(1-P_0)
\eeq

\noindent where $P_2$ is given exactly  by (\ref{eq3.6}).  This equation is 
easily solved as

\beq
\label{eq3.10}
1-P_0(y) \, = \, \exp\left[-{y\over{\sqrt{2\pi}}}\left({\alpha^3\over 
\beta}\right)^{1/2}e^{-\alpha/2\beta}\right].
\eeq

\noindent This is exactly the tunneling probability found by 
Alessandrini et al.~\cite{0dimq1,0dimq,Amati:1976ck,Ciafaloni:1977xv,0dimi} 
and given in (\ref{Q14}) above.

\indent The factor multiplying $-y$ in the exponent in (\ref{eq3.10}) is to 
be identified with $\alpha_P-1$ with $\alpha_P$ the pomeron 
intercept.  The high energy scattering amplitude will decrease at 
large  $y$ a result here which is easily interpreted as the loss of 
probability from the active states, $n\not= 0,$ to the inactive, 
$n=0$, state.  In the reaction diffusion picture there is no 
tunneling, only a very slow loss of probability from the metastable 
fixed point configuration of probabilities given by (\ref{eq3.6}) into the 
$n=0$ state.

\subsection{ $3P$ and $4P$ couplings}

The hierarchy equation (see (\ref{mark1}))

\beq\label{eq3.11}
{dn^{(k)}\over dy} \, = \, \alpha k n^{(k)}+ \alpha k (k-1) 
n^{(k-1)}-\beta k n^{(k+1)}-\beta k (k-1) n^{(k)}
\eeq

\noindent along with the corresponding master equation

\beq\label{eq3.12}
{dP_n\over dy}\, =\, - \alpha nP_n + \alpha (n-1) P_{n-1} + \beta n 
(n+1) P_{n+1} - \beta n (n-1) P_n
\eeq

\noindent correspond to a genuine reaction diffusion process with a 
$1\to 2$ vertex of strength $\alpha$ and a $2\to 1$ vertex of 
strength $\beta$, see e.g.\ Ref.~\cite{stoch3}.   
In Reggeon field theory language the additional 
term in (\ref{eq3.11}), as compared to (\ref{eq3.1}), 
corresponds to a $4P$ coupling. The fixed point solution to 
(\ref{eq3.11}) and (\ref{eq3.12}) is

\beq\label{eq3.13}
P_n\, = \, {1\over n!} \left({\alpha\over \beta}\right)^n {1\over 
e^{\alpha/\beta}-1}
\eeq

\noindent and

\beq\label{eq3.14}
n^{(k)}\, = \, (\alpha/\beta)^n {e^{\alpha/\beta}\over 
e^{\alpha/\beta}-1}.
\eeq

\noindent Eq.~(\ref{eq3.12})
does not have transitions to the $n=0$ state so if 
one takes a normalized initial condition having $P_0=0$ then $P_0(y)$ 
will remain equal to zero.  When $\alpha/\beta \gg 1$ the fixed point 
solution is very close to a Poisson distribution of probabilities. 
The fact that there is now a genuine fixed point solution to the 
reaction-diffusion problem guarantees that $\alpha_P=1$ in the 
solution to the Reggeon field theory, and this indeed is the case as
we have seen in Sec.~2.5.

\subsection{ High energy amplitudes}

As discussed earlier in Sec.~2.3 
it is not so straightforward to decide how to define a high energy 
scattering amplitude.  In Reggeon field theories one has to decide 
how the pomeron field, the dynamical variable in the theory, couples 
to external particles which  are not part of the dynamics of the 
theory.  The most common choice, given in (\ref{ext1}) and (\ref{ext2}), 
has been to directly couple the pomeron field to a source at $y=0$ 
with strength, say, $g_1$ and to another source of strength, say, 
$g_2$ at $y=Y$. 
In the following we shall deal exclusively with the 
model described in Sec.~3.2 as this seems the more interesting case. 
The $3P$ case is easily treated by the same methods and the result is
as in (\ref{Q14}).

\indent Once an $s$-channel picture of the dynamics is introduced 
more constraints on the definition of the scattering emerge.  In the 
first place the particles which scatter are the same as the particles 
making up the $s$-channel dynamics.  Secondly, the scattering is no 
longer manifestly boost invariant.  For example, in the center of 
mass the scattering looks like a scattering of two highly evolved 
systems while in the rest frame of one of the particles the 
scattering looks like that of a highly evolved system on an 
elementary quantum.  For both (\ref{eq3.1}) and (\ref{eq3.11}) 
boost invariance can be satisfied by taking the scattering amplitude to be

\beq\label{eq3.15}
T(Y) \, = \, \sum_{k=1}^\infty {(-1)^{k-1}\over 
k!}\alpha^{2k}n^{(k)}(Y_0)n^{(k)}(Y-Y_0)
\eeq

\noindent where $Y_0$ is arbitrary.  Eq.~(\ref{eq3.15}) represents scattering 
as the sum over  $k$  interactions of elementary quanta in one, say 
left-moving, system evolved to rapidity $Y-Y_0$ and another, 
right-moving, system evolved to rapidity $Y_0.$ The elementary 
interactions are of strength  $\alpha^2$ and $T$ is independent of 
$Y_0$ if $\alpha^3=\beta.$  The use of factorial moments seems 
necessary in (\ref{eq3.15}) in order to get boost invariance.  At large  $y$ 
values  $n^{(k)}(y)$ approaches the fixed point solution given in 
(\ref{eq3.14}), independent of the initial condition, in the case of a 
genuine reaction diffusion process corresponding to $3P$ and $4P$ 
interactions.  $T$ is most easily evaluated by taking $Y_0=0$ in 
which case

\beq\label{eq3.16}
T(Y) \, = \, \alpha^2 n^{(1)}(Y)
\eeq

\noindent which has the large  $Y$ limit

\beq\label{eq3.17}
T(Y)\, {}_{\widetilde{Y\to \infty}} \, \alpha^3/\beta \, {1\over 
1-e^{\alpha/\beta}}\, \simeq \, 1+e^{-1/\alpha^2}
\eeq

\noindent where we have taken $\alpha^3=\beta$ in the right-most 
expression in (\ref{eq3.17}). This agrees with the result found from Reggeon field theory in (\ref{CVF11}). The fact that $T$ is very near one at large 
$Y$ and small $\alpha$ follows from the fixed point properties of the 
theory along with boost invariance and is independent of the initial 
condition for the system so long as the initial states are normalized 
to unit probability.

\indent Now let us evaluate the scattering amplitude in the eikonal 
model, repeating the derivation given in Sec.~2.5 but from a 
different perspective. It is still convenient to use reaction-diffusion 
dynamics, but we shall define $T$ in the way that has been traditional in 
Reggeon field theories.  That is, we take, corresponding to (\ref{ext2}),

\beq\label{eq3.18}
T(Y) \, = \, \sum_{k=1}^\infty {(-1)^{k-1}\over k!} \, (g_2\alpha)^k\, 
n^{(k)}(Y)
\eeq

\noindent where $n^{(k)}$ satisfies the hierarchy equations (\ref{eq3.11}) of 
the Reggeon field theory.  The initial condition for $n^{(k)}$ is

\beq\label{eq3.19}
n^{(k)}(0) \, = \, (g_1/\alpha)^k
\eeq

\noindent expressing the eikonal model of pomeron interactions with 
the external particle (source) at $y=0.$  Now (\ref{eq3.19}) requires

\beq\label{eq3.20}
P_n(0)\, = \, {1\over n!}(g_1/\alpha)^n e^{-g_1/\alpha}.
\eeq

\noindent We note that

\beq\label{eq3.21}
\sum_{n=1}^\infty P_n(0) \, = \, 1 - e^{-g_1/\alpha}
\eeq

\noindent so that the initial state is not properly normalized from 
the $s$-channel (reaction-diffusion) point of view.  Eq.~(\ref{eq3.21}) 
gives

\beq\label{eq3.22}
n^{(k)}(Y)\,{}_{\widetilde{Y\to\infty}}\,  
(1/\alpha^2)^k \,  (1-e^{-g_1/\alpha}) \,
{1 \over 1-e^{-1/\alpha^2} }
\eeq

\noindent which, when used in (\ref{eq3.18}), gives, as in (\ref{CVF9}), 

\beq\label{eq3.23}
T(Y)\, {}_{\widetilde{Y\to\infty}} \, 
(1-e^{-g_2/\alpha})(1-e^{-g_1/\alpha})\, {1\over 
1-e^{-1/\alpha^2}}
\eeq

\noindent or

\beq\label{eq3.24}
T(Y)\, {}_{\widetilde{Y\to\infty}}\,
(1-e^{-g_2/\alpha})(1-e^{-g_1/\alpha})
\eeq

\noindent where we have dropped small terms, of size 
$e^{-1/\alpha^2},$ in obtaining (\ref{eq3.24}) which is the same as given in 
Refs.~\cite{Amati:1976ck,Ciafaloni:1977xv,bm}.

\indent Eq.~(\ref{eq3.24}) expresses ``gray'' rather than ``black'' 
scattering.  From the Reggeon field theory point of view this result 
is rather mysterious; why should a strongly interacting scattering 
not saturate unitarity at high energy?  From the reaction-diffusion 
($s$-channel) picture the factors in (\ref{eq3.24}) just represent the 
normalization of the states (see (\ref{eq3.21})) involved in the scattering.

\subsection{Generating functions}

For completeness, let us finish with a brief discussion of a generating 
function, $Z(y,v)$, for occupation number probabilities, $P_n(y)$, 
in the reaction diffusion process, 
\beq
Z(y,v) \, = \, \sum_{n=1} ^{\infty} P_n(y)\, v^n,  
\eeq
where $v$ is an auxiliary variable.
The probabilities of $n$-particle states may be expressed as
\beq
P_n(y) \, = \, {1\over n!} \,
\left. {\partial^n Z(y,v) \over \partial v^n} \right|_{v=0},
\eeq
and the factorial moments read
\beq
\label{fact}
n^{(k)}(y) \, = \, \left. {\partial^k Z(y,v) \over \partial v^k} \right|_{v=1}.
\eeq
The generating function is intimately connected to the wave function
of the corresponding RFT, given by (\ref{Quant14}). 
Using (\ref{npsi}), (\ref{Quant14}), 
the identification $\,\langle u^k \rangle_y = n^{(k)}(y)\,$ 
and the relation in (\ref{fact}) we obtain
%
\beq
\Psi(y,q)
\, = \,
- \sum_{k=1} ^{\infty}  \, {1\over k!} \left. {\partial^k Z(y,v) \over \partial v^k} \right|_{v=1}\, (-\alpha q)^k, 
\eeq
which may be simplified to
\beq
\Psi(y,q) 
\, = \,
Z(y,1)-Z(y,1-\alpha q).
\eeq
It is straightforward to verify that the inverse relation reads
\beq
Z(y,v) \, = \,  \Psi(y,1/\alpha) - \Psi(y,(1-v)/\alpha).
\eeq


\section*{Acknowledgments}
We are grateful to J.~Bartels, C.~Ewerz, E.~Iancu, S.~Munier, 
M.~Salvadore and G.P.~Vacca for enlightening discussions.
A.M.\ and A.~Sh.\ wish to thank the DESY theory group for their hospitality 
during their visit when this work was being initialized. 
S.B.\ thanks the Minerva foundation for its support,
L.M.\ gratefully acknowledges the support of the grant 
of the Polish State Committee for Scientific Research 
No.\ 1~P03B~028~28. A.M.\ and B.X.\ are partially supported by the
US~Department of Energy. A.~Sh. acknowledges financial support by the 
Deutsche Forschungsgemeinschaft under contract Sh~92/2--1.

\end{document}